\documentclass[]{interact}
\usepackage[utf8]{inputenc}
\usepackage[T1]{fontenc}
\usepackage{natbib}
\usepackage[english]{babel}
\usepackage{array}
\usepackage{multirow}
\usepackage{amsmath}
\usepackage{amssymb}
\usepackage{enumitem}
\usepackage{graphicx}
\usepackage{subfigure}
\graphicspath{{figures/}}
\DeclareGraphicsExtensions{.pdf,.png,.jpg}
\usepackage[unicode]{hyperref}
\theoremstyle{plain}

\theoremstyle{definition}

\theoremstyle{remark}

\begin{document}


\title{Early indications of anomalous behavior in the 2019 spring ozone hole over Antarctica}

\author{
\name{Gennadi Milinevsky\textsuperscript{a,b,c}\thanks{CONTACT Gennadi Milinevsky. Email: genmilinevsky@gmail.com}, Oleksandr Evtushevsky\textsuperscript{b}, Andrew Klekociuk\textsuperscript{d,e}, Yuke Wang\textsuperscript{a}, Asen Grytsai\textsuperscript{b}, Valerii Shulga\textsuperscript{a,f}, Oksana Ivaniha\textsuperscript{b} }
\affil{\textsuperscript{a}College of Physics, International Center of Future Science, Jilin University, Changchun, 130012, China;\\ \textsuperscript{b}Taras Shevchenko National University of Kyiv, Kyiv, 01601, Ukraine;\\ \textsuperscript{c}National Antarctic Scientific Center, MES of Ukraine, Kyiv, 01601, Ukraine;\\ \textsuperscript{d}Antarctica and the Global System, Australian Antarctic Division, Kingston, 7050, Australia;\\ \textsuperscript{e}School of Earth Sciences, University of Melbourne, Melbourne, 3053, Australia;\\ \textsuperscript{f}Institute of Radio Astronomy, NAS of Ukraine, Kharkiv, 61002, Ukraine}
}

\maketitle

\begin{abstract}
	
The level of quasi-stationary planetary wave (QSW) activity in the Antarctic winter stratosphere provides insights into the likely behavior of the ozone hole in the following spring months. Observation of an anomalously large amplitude of the QSW in winter stratospheric temperatures is an indicator that strong disturbances to the polar vortex are likely to occur, and may lead to large reductions in both the area of the Antarctic ozone hole area and the overall amount of stratospheric ozone that is depleted. In the sudden stratospheric warming (SSW) preconditions in 2019, the maximum QSW amplitude over Antarctica in August was approximately 12 K, which was only 2 K less than conditions prior to the unprecedented major Antarctic SSW in 2002. Under these conditions, the Antarctic SSW in 2019 has the potential to become a major SSW. The additional factors disturbing the Antarctic stratosphere in 2019 was anomalously warm sea surface temperatures (SSTs) in the central tropical Pacific Ocean and western Indian Ocean, and the descending easterly phase of the Quasi-Biennial Oscillation. The combination of these factors – the large amplitude of the QSW, the warm tropical SSTs and transitioning phase of the QBO – has the potential to cause the early disruption of the ozone hole and reduce the overall level of ozone depletion in 2019, and may also have important regional consequences for weather conditions in the Southern Hemisphere.
\end{abstract}


\section*{Introduction}

Although the ozone hole over Antarctic develops in the austral spring months of September–November, it is strongly influenced by the state of the Antarctic stratosphere in the preceding winter (June–August). Dynamical disturbance of the winter stratospheric vortex by planetary waves weakens the strength of the vortex and reduces the amount of ozone that is depleted within the ozone hole during spring \citep{Shindell1997,Allen2003,Huck2005,Grassi2008,Weber2011}. As known, due to the unstable atmosphere over the Arctic, such events as SSW occur almost every second year \citep{Charlton2007}. Observations of quasi-stationary planetary waves (QSW) activity in the lower stratosphere during August provides insight into the state and dynamics of the ozone hole that will take place in the following spring months. As discussed in \citep{Grytsai2008,Kravchenko2012,Evtushevsky2019b}, the amplitude of the QSW in stratospheric temperature serves as a key predictor of both the level of dynamical disturbance that the polar vortex is likely to display in spring and the final breakdown date of the ozone hole. 

An additional factor that potentially influences the level of disturbance to the ozone hole is the state of tropical sea surface temperatures (SSTs). In \citep{Evtushevsky2015,Evtushevsky2019a} was shown that SST anomalies in the central Pacific in June can force disturbances in the temperature of the Antarctic stratosphere which have peak influence in the following October. This link arises when wave trains generated by deep convection over warm SST anomalies propagate poleward from the troposphere to the Antarctic stratosphere \citep{Grassi2008,McIntosh2017}. In 2019, the combination of the quasi-stationary wave 1 (QSW-1) in extratropical stratospheric temperatures having a large amplitude in August and strong surface warming in the central tropical Pacific Ocean, as well as in the western Indian Ocean, in June has the potential to cause an early disruption of the ozone hole. 

The ozone hole evolution is also influenced by the Quasi-Biannual Oscillation (QBO) due to modulation of the polar vortex strength \citep{Holton1980,Watson2014}. The SSW appears earlier and is more intense, and planetary wave amplitude is larger for the easterly QBO phase \citep{Holton1991,Anstey2010}. In this paper, based on a SSW predictor analysis, the stratospheric preconditioning that took place in the Antarctic austral winter of 2019 is discussed. 

\section*{Data and method}
Similar to \citet{Kravchenko2012} we use the amplitude of QSW in August at $50-80^{\circ}$S and 50 hPa to provide an indication of October ozone hole area. The amplitude of the wave is obtained from a zonal monthly average temperature data from the NCEP–NCAR reanalysis (NNR; \url{https://www.esrl.noaa.gov/psd/data/gridded/data.ncep.reanalysis.html}). We use also NRR data to obtain the June monthly average SST for the central tropical Pacific ($20^{\circ}$N– $20^{\circ}$S, $160 -220^{\circ}$E), as was done by \citet{Kravchenko2012}. HadISST data for the tropical SSTs were also used from \url{http://hadobs.metoffice.gov.uk/hadisst/}.
A additional atmospheric parameters are illustrated by data from the Climate Prediction Center (\url{https://www.cpc.ncep.noaa.gov/products/stratosphere/polar/polar.shtml}), NASA Ozone Watch (\url{https://ozonewatch.gsfc.nasa.gov/}), NASA GSFC Atmospheric Chemistry and Dynamics Laboratory (\url{https://acd-ext.gsfc.nasa.gov/Data_services/met/qbo/}) and the KNMI Climate Explorer (\url{https://climexp.knmi.nl}).
\section*{Results and discussion}

\begin{figure}[!h]
	\center{\includegraphics[scale=0.9]{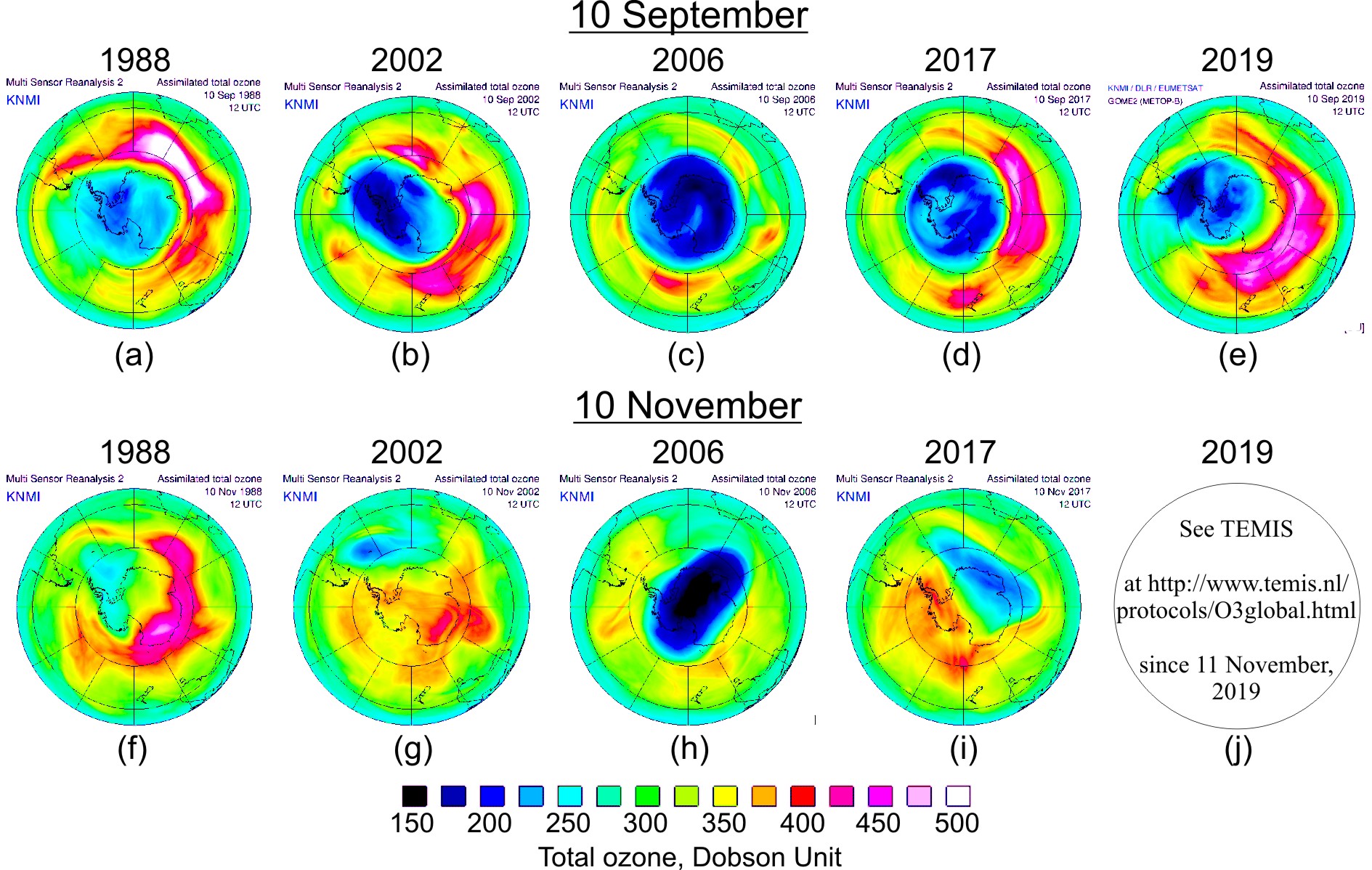}}
		\caption{Ozone hole over Antarctic in September 10: (a) 1988, (b) 2002, (c) 2006, (d) 2017 and (e) 2019, and in November 10: (f) 1988, (g) 2002, (h) 2006 and (i) 2017 according to TEMIS Multi Sensor Reanalysis (except (e) with METOP-B satellite data for 2019 using GOME2 spectrometer, \url{http://www.temis.nl/}).}
		\label{ris:fig1}
\end{figure}

As discussed in \citep{Evtushevsky2019b}, QSW-1 had an important role in disturbing the Antarctic polar vortex in 1988, 2002 and 2017. As shown in the Figure \ref{ris:fig1}, the total column ozone distribution over the Southern Hemisphere in the early part of spring in 2002 and 2019 exhibited strong zonal asymmetry. 

A relatively small ozone hole and high values in the mid-latitudes of the eastern hemisphere are observed (Figure \ref{ris:fig1}b and \ref{ris:fig1}e) under the influence of the meridional transport associated with the quasi- stationary wave. 

\begin{figure}[!h]
	\center{\includegraphics[scale=1.1]{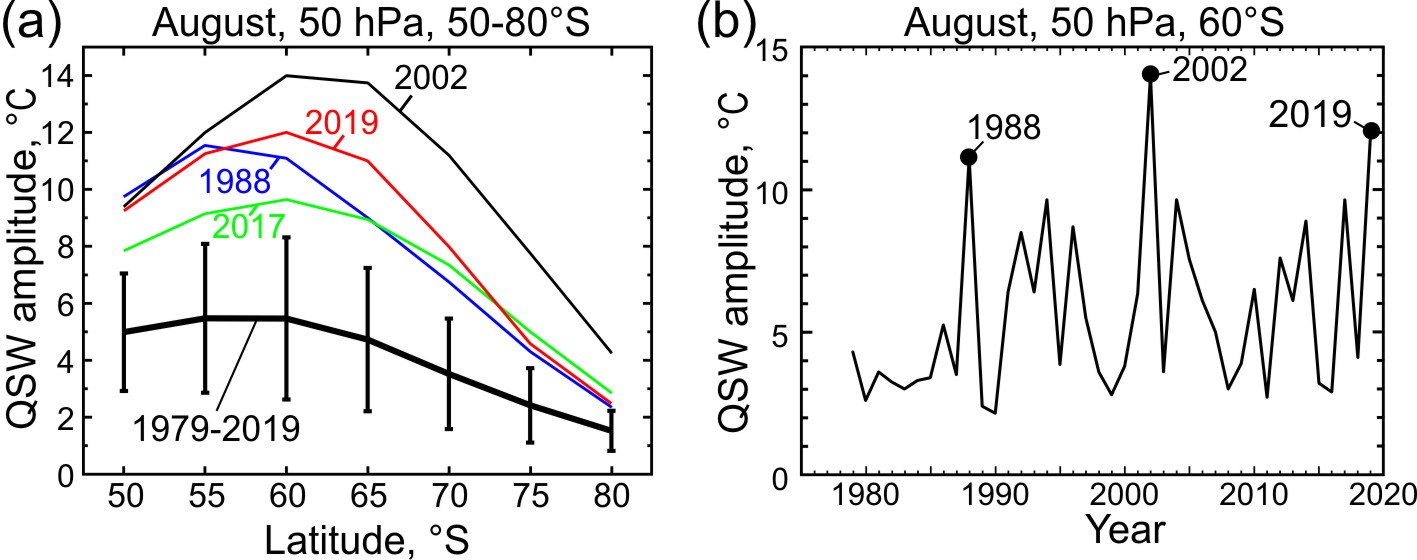}}
	\caption{(a) Latitudinal QSW-1 amplitude variation at 20 km (50 hPa) in August for the four anomalous years 1988, 2002, 2017 and 2019 versus the 1979–2017 climatology (thick curve with the standard deviation shown by vertical bars); (b) time series of the QSW-1 amplitude in the Antarctic stratosphere in August at the $60^{\circ}$S latitude circle.}
	\label{ris:fig2}
\end{figure}

In the late spring, the ozone hole had collapsed much earlier (Figure \ref{ris:fig1}g) in 2002 than more typically observed (usually late November to early December; \url{https://ozonewatch.gsfc.nasa.gov/meteorology/SH.html}). In 2019, the ozone distribution in early September (Figure \ref{ris:fig1}e) was similar to the other disturbed years, except that the mid-latitude ‘ozone collar’ region was relatively strong and somewhat similar to 2002 (Figure \ref{ris:fig1}b).

\begin{figure}[!h]
	\center{\includegraphics[scale=1.1]{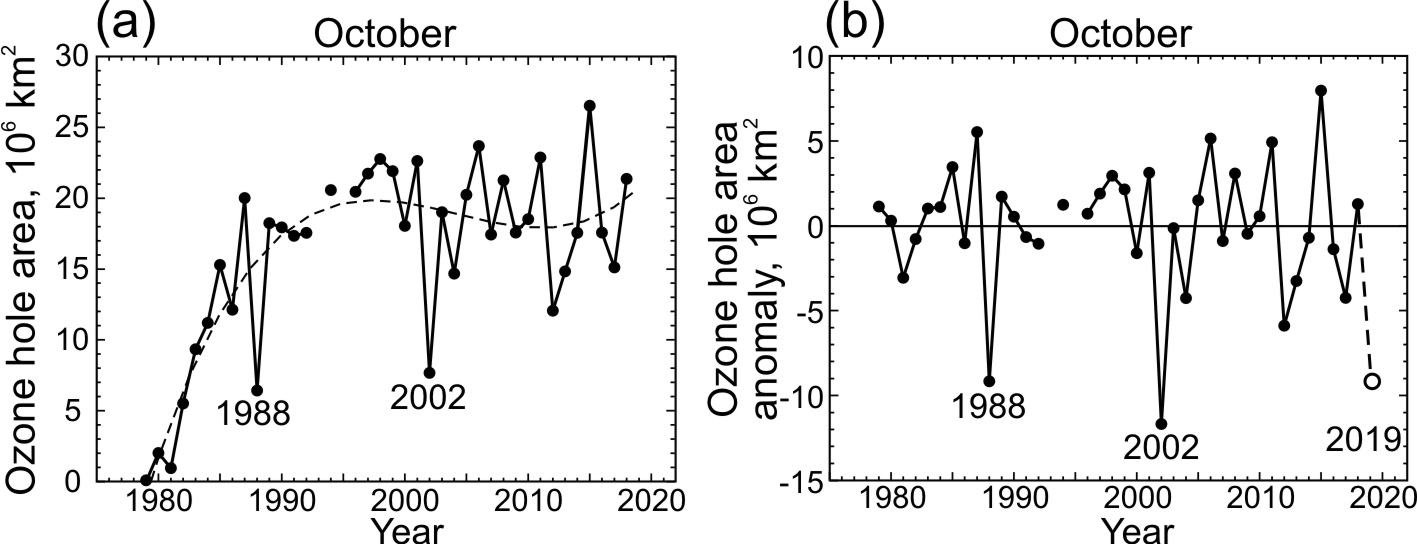}}
	\caption{(a) October averaged ozone hole area over 1979–2018, dashed curve shows fitting with the third degree polynomial; (b) the ozone hole area anomaly in October from the detrended time series (a) with the 3rd polynomial removed and with the predicted level for 2019 (open circle).}
	\label{ris:fig3}
\end{figure}

The high QSW amplitude in August 2019 (Figure \ref{ris:fig2}a (red curve) and Figure 2b) was similar to the precondition in 2002 and suggests the possibility of a significant reduction of the area of the ozone hole in the forthcoming spring. Based on the QSW amplitude level in August 2019 (Figure \ref{ris:fig2}), the possible ozone hole area anomaly relative to the mean tendency (dashed curve in Figure \ref{ris:fig3}a) in October 2019 could be between that in 1988 and 2002 (open circle in Figure \ref{ris:fig3}b).

\begin{figure}[!h]
	\center{\includegraphics[scale=0.9]{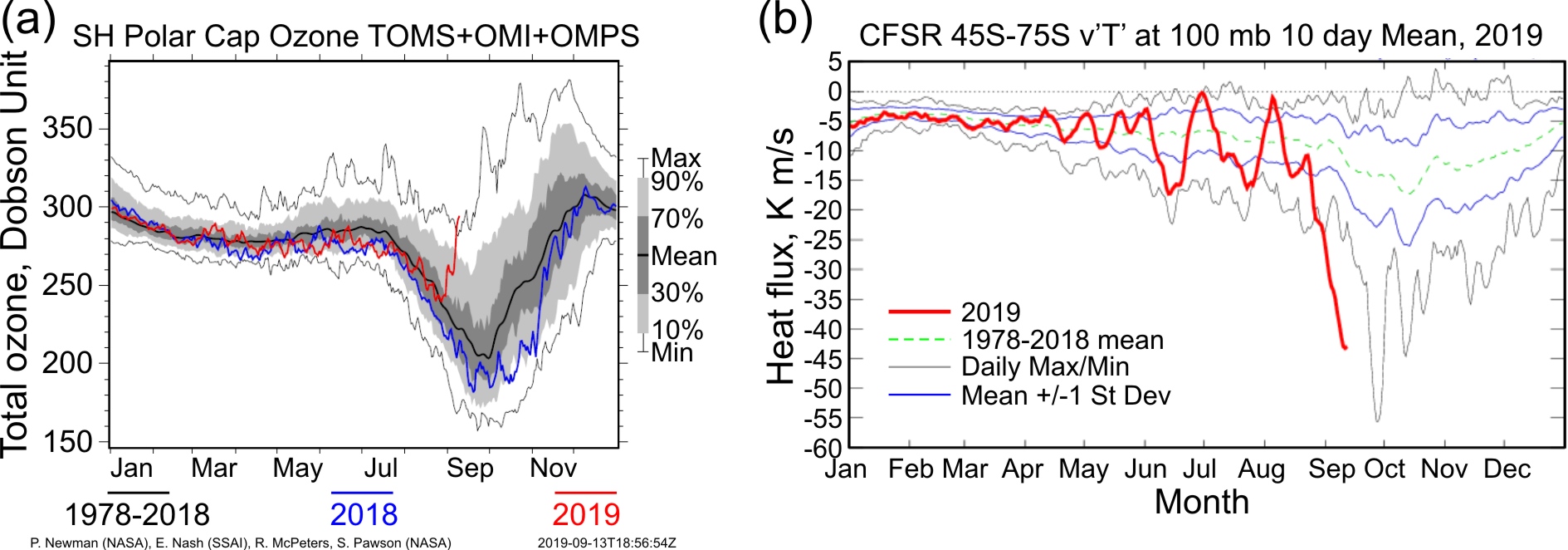}}
	\caption{(a) The total column ozone averaged around the polar cap for latitudes south of $63^{\circ}$S (\url{https://ozonewatch.gsfc.nasa.gov/meteorology/SH.html}) and (b) the 10 day averaged eddy heat flux towards the South Pole (\url{https://www.cpc.ncep.noaa.gov/products/stratosphere/polar/polar.shtml}) estimated at the 100 hPa pressure level. Red curves are for January–September 2019.}
	\label{ris:fig4}
\end{figure}
	
In early September 2019, the ozone hole area was reduced to $12-16\times 10^6$~km$^2$ as seen from the NASA Ozone Watch data (\url{https://ozonewatch.gsfc.nasa.gov/meteorology/SH.html}, last accessed 14 September 2019). For comparison, the ozone hole area during last two decades varied in the range $20-25\times 10^6$~km$^2$ (see Figure \ref{ris:fig3}a) at a similar time of year. The total column ozone averaged around the polar cap for latitudes south of $63^{\circ}$S is shown in Figure \ref{ris:fig4}a, where red line shows the lowest values of total ozone in September 2019 in comparison to a range of variability in September 1979–2018. This development of stratospheric processes indicates a possible major SSW in September–October, similar to that observed in September 2002 \citep{Allen2003} which has occurred only once during observations of the ozone hole.

Strong warming in the Antarctic stratosphere in the early spring of 2019 is associated with anomalous poleward heat flux (red curve in Figure \ref{ris:fig4}b). Strong negative fluxes indicate poleward transport of heat via eddies that may result in a smaller polar vortex and an earlier transition from winter to summer circulations (\url{https://www.cpc.ncep.noaa.gov/products/stratosphere/polar/polar.shtml}), \citep{Limpasuvan2005}.

\begin{figure}[!h]
	\center{\includegraphics[scale=0.9]{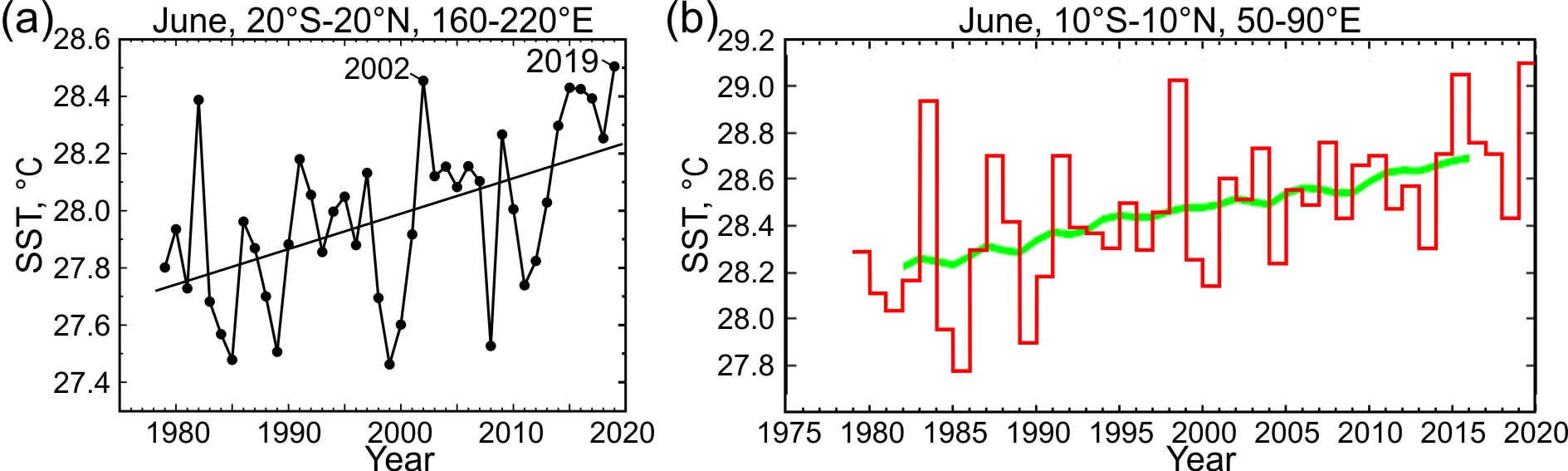}}
	\caption{The mean June sea surface temperature (SST) in (a) the central Pacific Ocean ($20^{\circ}$S-$20^{\circ}$N, $160-220^{\circ}$E) and (b) the western Indian Ocean ($10^{\circ}$S-$10^{\circ}$N, $50-90^{\circ}$E). In 2019, the temperatures in these regions were the highest in 40 years of observations $+28.5^{\circ}$C (central Pacific Ocean) and $+29.1^{\circ}$C (western Indian Ocean).}
	\label{ris:fig5}
\end{figure}

The second additional index to predict the future behavior of the ozone hole this year is the index of the sea surface temperature (SST) variability in the central tropical Pacific \citep{Evtushevsky2015,Evtushevsky2019a}. In June 2019, the SST in this region increased to $+28.5^{\circ}$C, the highest in the last 4 decades (Figure \ref{ris:fig5}a).

\begin{figure}[!h]
	\center{\includegraphics[scale=0.9]{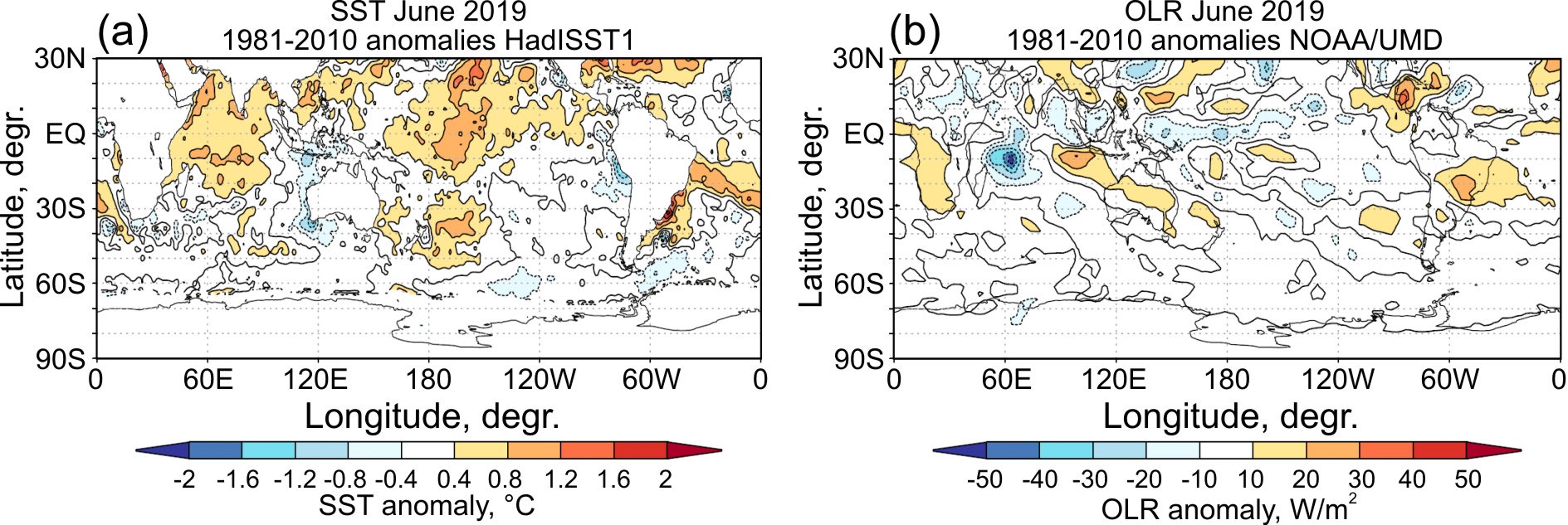}}
	\caption{(a) Mean SST anomalies for June 2019 from the HadISST reconstruction. (b) Mean OLR anomalies in June 2019 from the NOAA/UMD OLD database. The base period is 1981-2010.}
	\label{ris:fig6}
\end{figure}

Tropical convection perturbations caused by tropical temperature anomalies, as known by the El Nino phenomena, have a global impact \citep{Domeisen2019}. The western Indian Ocean SSTs ($10^{\circ}$S to $10^{\circ}$N, $50^{\circ}$E to $90^{\circ}$E) were also at anomalous high levels for June ($+29.1^{\circ}$C, Figure \ref{ris:fig5}b). The warm SST anomalies (Figure \ref{ris:fig6}a) are associated with a strong negative anomaly in outgoing long wave radiation (OLR; Figure \ref{ris:fig6}b). 

\begin{figure}[!h]
	\center{\includegraphics[scale=0.9]{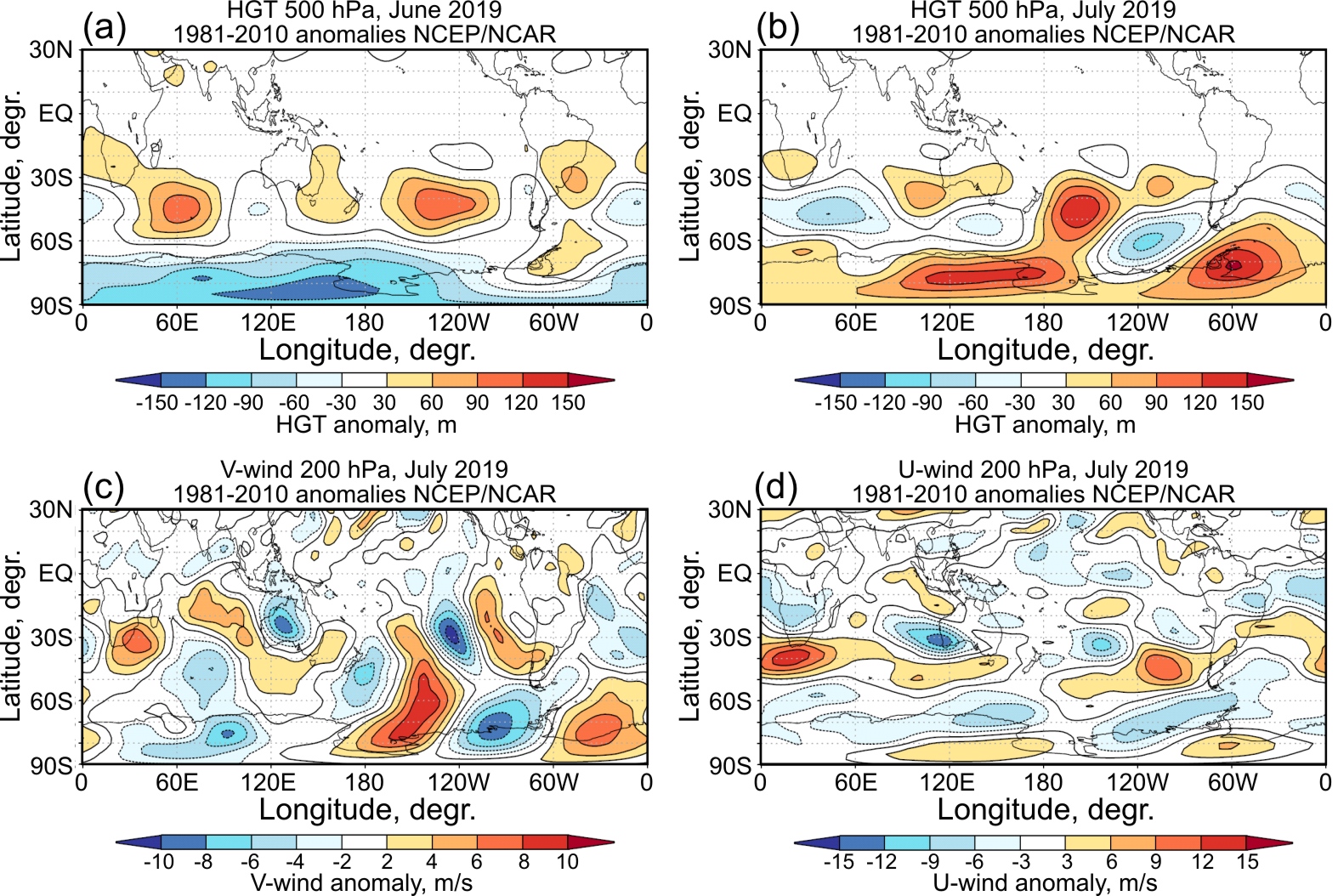}}
	\caption{(a) Mean SST anomalies for June 2019 from the HadISST reconstruction. (b) Mean OLR anomalies in June 2019 from the NOAA/UMD OLD database. The base period is 1981-2010.}
	\label{ris:fig7}
\end{figure}

The negative OLR anomalies in the Pacific are not as strong, but more distributed in longitude which is also an important factor. Strong negative OLR anomalies are associated with increased deep convection and enhanced Rossby wave activity \citep{McIntosh2017}. The SST anomalies in both the Pacific and Indian oceans appear to generate poleward propagating wave trains seen in zonal and meridional winds at 200 hPa and geopotential height at 500 hPa in July (Figure \ref{ris:fig7}). The simultaneous poleward wave driving from both the Pacific and Indian Ocean regions may be the unique aspect of the SSW 2019.

Our study shows that the June tropical disturbance through the stratospheric circulation branch reaches the Antarctic stratosphere and creates the greatest impact on its temperature after 4 months, in October \citep{Evtushevsky2015,Evtushevsky2019a}. The combination of the two factors, high amplitude of the temperature QSW in August and high tropical SST in June can cause the ozone hole to disappear early this year, at least in October, while usually a decay happened in November and December, as noted above.

\begin{figure}[!h]
	\center{\includegraphics[scale=0.8]{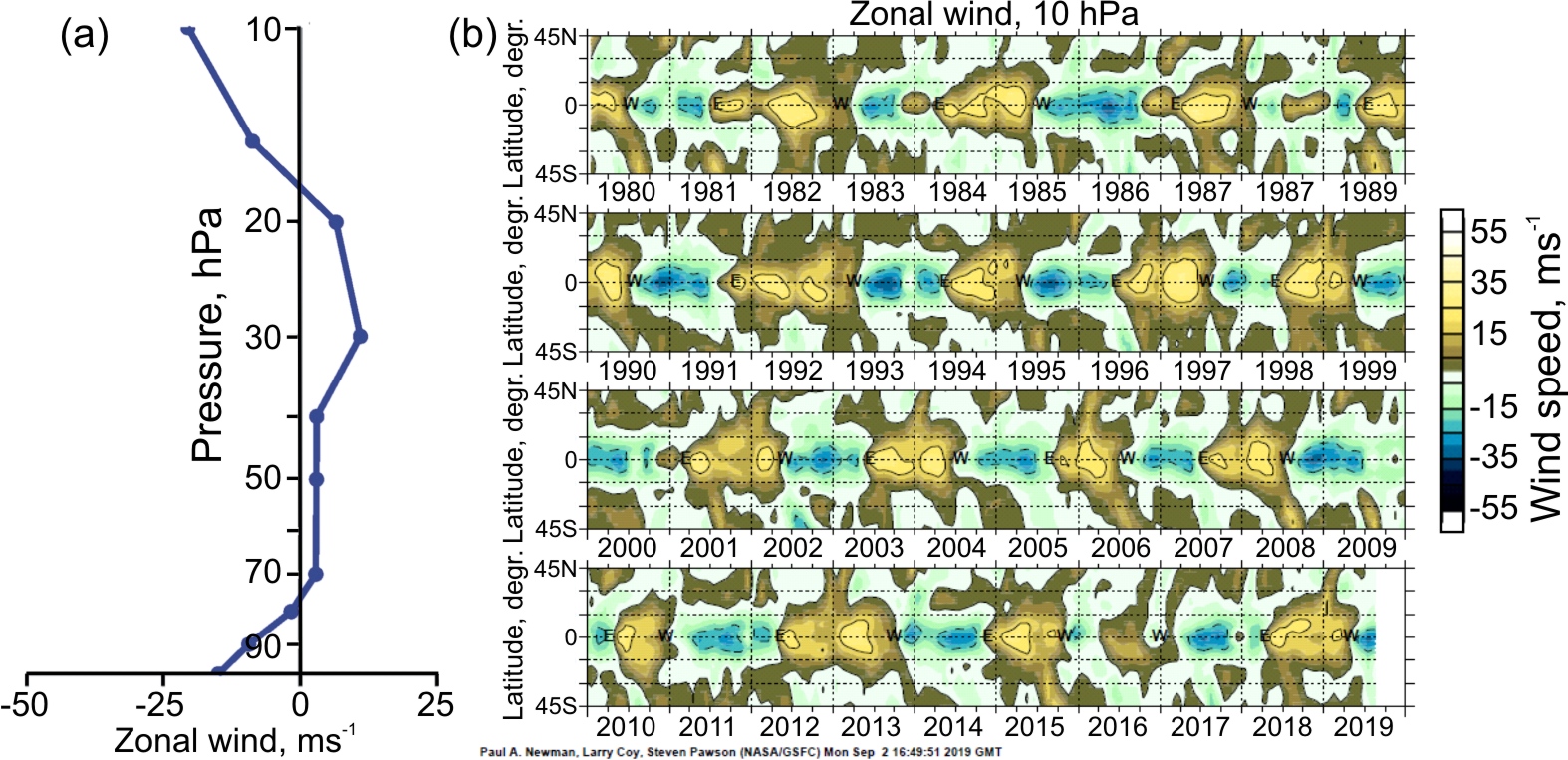}}
	\caption{ (a) The QBO in the Eastern Descending phase on September 9, 2019 (Singapore RAOB zonal wind). (b) Time-latitude cross section of the 10 hPa wind speed from the MERRA-2 reanalysis. From the NASA GSFC Atmospheric Chemistry and Dynamics Laboratory data (\url{https://acd-ext.gsfc.nasa.gov/Data_services/met/qbo/}).}
	\label{ris:fig8}
\end{figure}

Favorable factor for an unstable polar vortex over Antarctica in spring 2019 is also evolution of the quasi-biannual oscillation (QBO), which is in the easterly descending phase (Figure \ref{ris:fig8}). As noted in the Introduction above, the polar vortex and ozone hole disappear earlier in the easterly QBO phase \citep{Holton1980,Holton1991,Anstey2010,Watson2014}. The QBO wind at 10 hPa was easterly from May 2019 (Figure \ref{ris:fig8}b) and contributed to the polar vortex destabilization. Thus, the QBO acts in winter–spring 2019 in the same direction (to earlier SSW occurrence) as the stratospheric QSW in August and tropical SST anomalies in June.

\section*{Conclusions}

In this study, we have demonstrated an influence of the winter preconditioning on the spring ozone hole over Antarctica and the ability to predict the possible ozone hole area and ozone depletion level in austral spring 2019. The two proposed predictors, the quasi-stationary wave amplitude in the SH polar stratospheric temperature in August and sea surface temperature in the tropical Pacific and Indian Oceans in June were used. They indicate that ozone hole size in 2019 may be between those in 1988 and 2002, the years of the historically lowest ozone loss in the Antarctic spring. 

Along with the descending eastern phase of the QBO, this may precede the major SSW in September–October 2019, the second one since the unprecedented event in 2002. As distinct from the major SSW 2002 in Antarctica, which was classified as vortex split event due wave-2 dominance, the SSW 2019 is expected to be the vortex displacement event under the wave-1 influence (Figure \ref{ris:fig1}e). Over the Arctic, due to the unstable atmosphere, such events as major SSWs occur every second year and result in strong regional change in the regional surface climate. The recent event of a major SSW over the Arctic was observed in February 2018 and brought a cold outbreak to Canada, the US, and Ukraine in March \citep{Karpechko2018,Vargin2019,Wang2019}. The SSW over the Antarctic in 2019 may potentially bring warm and dry conditions to parts of eastern Australian and northern New Zealand \citep{Lim2018}.

\section*{Data availability}
The authors declare that all data which supports the findings are provided with the paper. All data is available from public sources. The sources of data are the NCEP–NCAR reanalysis (NNR; \url{https://www.esrl.noaa.gov/psd/data/gridded/data.ncep.reanalysis.html}). The atmospheric parameters are available from NASA Ozone Watch (\url{https://ozonewatch.gsfc.nasa.gov/}), the Climate Prediction Center (\url{https://www.cpc.ncep.noaa.gov/products/stratosphere/polar/polar.shtml}), NASA GSFC Atmospheric Chemistry and Dynamics Laboratory (\url{https://acd-ext.gsfc.nasa.gov/Data_services/met/qbo/}) and the KNMI Climate Explorer (\url{https://climexp.knmi.nl}). HadISST data for the tropical SSTs were used from \url{http://hadobs.metoffice.gov.uk/hadisst/}. Results of QSW and SST analysis are available from the corresponding author upon request.

\section*{Author contributions} 
GM proposed, coordinated and led the efforts for this manuscript. OE, AK, GM and AG analyzed data, discussed the results, and provided interpretation. GM, OE, AK, VS, YW and OI contributed to writing the manuscript.

\section*{Acknowledgments} 
This work was supported in part by Taras Shevchenko National University of Kyiv, project 19BF051-08; by the College of Physics, International Center of Future Science, Jilin University, China. We also thank the NOAA for the NCEP–NCAR reanalysis data and atmospheric parameters from NASA GSFC Atmospheric Chemistry and Dynamics Laboratory team, Met Office Hadley Centre team for HadISST data, the KNMI for the Climate Explorer research tool. 
\bibliographystyle{ascelike-new}
\bibliography{biblio}   
\end{document}